\documentclass[aps,prl,twocolumn,showpacs,superscriptaddress]{revtex4-2}

\usepackage{graphicx}
\usepackage{dcolumn}
\usepackage{bm}
\usepackage{braket}
\usepackage[most]{tcolorbox}

\begin{document}


\title{Directional Transport in Rydberg Atom Arrays via Kinetic Constraints and Temporal Modulation}


%

\author{Yupeng Wang}%
\affiliation{ 
Department of Physics and Astronomy, Purdue University, West Lafayette, IN 47907, USA
}%

\author{Junjie Wang}
\affiliation{ 
Department of Physics and Astronomy, Purdue University, West Lafayette, IN 47907, USA
}
\affiliation{ 
School of the Gifted Young, University of Science and Technology of China, Hefei 230026, China
}
\author{Aishik Panja}
\affiliation{ 
Department of Physics and Astronomy, Purdue University, West Lafayette, IN 47907, USA
}
\author{Xinghan Wang}
\affiliation{ 
Department of Physics and Astronomy, Purdue University, West Lafayette, IN 47907, USA
}%
\author{Qi-Yu Liang}
\affiliation{ 
Department of Physics and Astronomy, Purdue University, West Lafayette, IN 47907, USA
}%
\affiliation{Purdue Quantum Science and Engineering Institute, Purdue University, West Lafayette, IN 47907, USA}


\date{\today}

\begin{abstract}
We propose an experimentally feasible scheme to achieve directional transport of Rydberg excitations and entangled states in atomic arrays with unequal spacings. By leveraging distance-dependent Rydberg-Rydberg interactions and temporally modulated laser detunings, our method directs excitation flow without requiring local addressing. Numerical simulations demonstrate robust and coherent transport under experimentally realistic conditions. Additionally, we show that this scheme enables controlled transport of Bell pairs and preserves entanglement during propagation. The approach provides a versatile platform for programmable directional transport, with potential applications in quantum simulation, entanglement distribution, and the design of scalable quantum processors and networks.

\end{abstract}

\maketitle


\textit{Introduction.}--Chirality or nonreciprocity, arising from broken spatial inversion or time-reversal symmetries, underpins a wide range of quantum phenomena. These include the emergence of unconventional topological phases, such as chiral spin liquids and Floquet-engineered anomalous topological systems~\cite{claassen2017dynamical,sun2023engineering,wintersperger2020realization}, directional light-matter interactions in chiral quantum optics~\cite{lodahl2017chiral}, and chiral transport, characterized by unidirectional excitation flow in various quantum materials~\cite{mcdonald2018phase,guo2022switchable}.
These phenomena are not only of fundamental interest but also essential for advancing quantum technologies such as programmable quantum simulations of lattice gauge theories~\cite{li2022coherent,banuls2020simulating}, efficient distributed quantum computing and networking~\cite{almanakly2024deterministic,grankin2018free}, and backaction-immune quantum sensing~\cite{wang2024quantum,xie2024quantum}. Specifically, programmable directional transport is pivotal for realizing dynamic control of quantum information flow, enabling novel architectures for quantum computation, simulation, and sensing.
However, achieving such transport remains an open challenge.

Here, we propose a novel approach to control the flow of Rydberg excitations in atomic arrays with unequal spacings. The distance-dependent Rydberg-Rydberg interactions induce local energy shifts, making an atom’s excitation energy dependent on nearby Rydberg excitations. By dynamically adjusting the global laser detuning, we offset these shifts and selectively steer excitations along predefined pathways. This approach enables laser-guided transport of quantum information encoded in ground-Rydberg qubits between spatially separated zones. The specific encoding within each zone may differ, provided it can be efficiently mapped to the transport encoding.
Unlike prior Rydberg-based directional transport protocols~\cite{li2022coherent,valencia2024rydberg, kitson2024rydberg}, our method eliminates the need for local addressing, simplifying experimental implementation while retaining deterministic control over quantum information relocation.


We focus on a 1D geometry with alternating spacings to numerically study the transport of a single excitation or a Bell pair. 
By leveraging the interplay between staggered spacing and modulated driving fields, we demonstrate directional and coherent transport under experimentally relevant conditions. We evaluate the impact of decoherence and position disorder on excitation transfer and entanglement preservation. These results pave the way for realizing controlled excitation transport in the state-of-the-art experimental platforms.

\textit{Model.}--We operate in the ``facilitated" or ``antiblockade" regime~\cite{ates2007antiblockade,amthor2010evidence,marcuzzi2017facilitation, zhao2024observation, liu2022localization, nill2024avalanche,magoni2023molecular, magoni2024coherent, wintermantel2021epidemic, helmrich2020signatures, fleischhauerpercolation, urvoy2015strongly,  ding2024ergodicity,wu2024nonlinearity, liu2024microwave}:  The ground-to-Rydberg transition is driven off-resonantly. The kinetic constraint ensures that resonance is achieved if and only if one of the nearest neighbors is in the Rydberg state. A laser beam with Rabi frequency $\Omega$ couples the ground ($\ket{0}$) and Rydberg ($\ket{1}$) states. 

\begin{figure}[ht]
\centering
\includegraphics[width=0.48\textwidth]{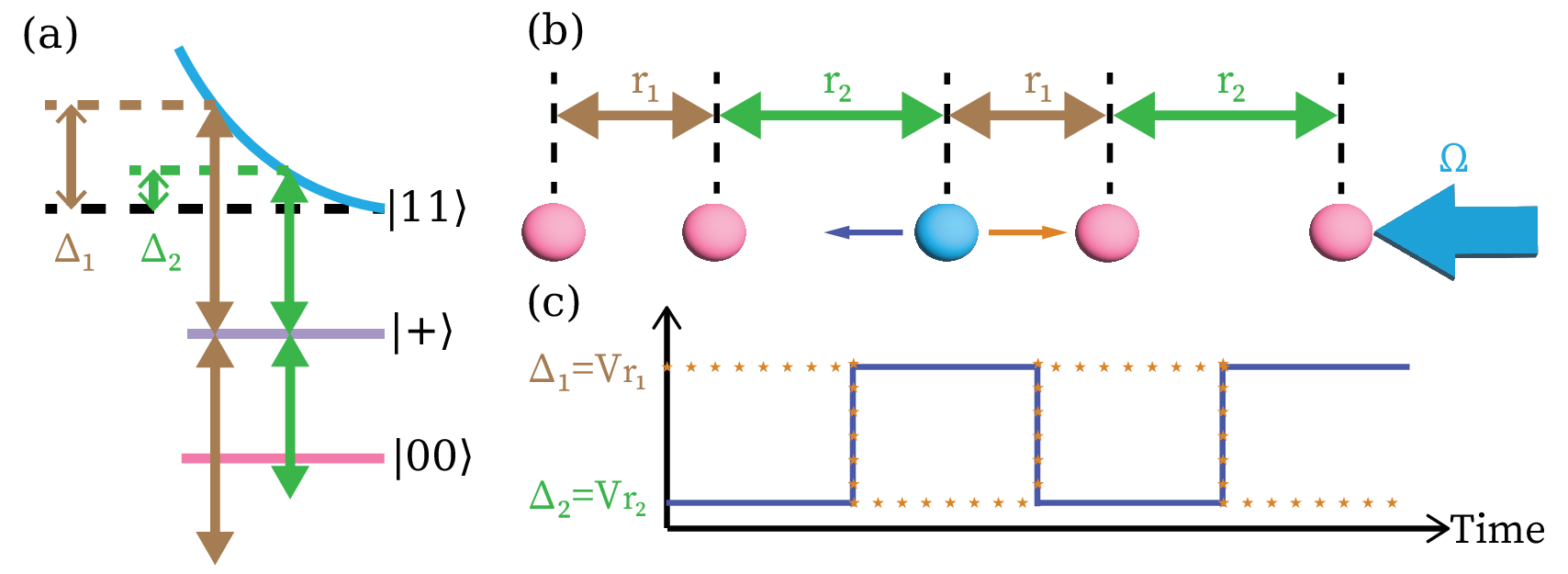}
\caption{Our excitation transport scheme. (a) The van der Waals interaction shifts the two-excitation state $\ket{11}$, with the interaction strength $V_{r_i}$ depending on the interatomic distance $r_i$ for $i=1,2$. To resonantly excite the second Rydberg excitation, the detuning relative to the $\ket{0}\leftrightarrow\ket{1}$ transition frequency is $\Delta_i=V_{r_i}$. (b) A Rydberg excitation (blue sphere) can be driven towards the left (right) by applying sequential $\pi$ pulses with alternating detunings starting from $\Delta_2$ ($\Delta_1$). The corresponding pulse sequences for leftward (rightward) transport are shown in (c) as solid blue (dashed orange) lines.}
\label{fig:array}
\end{figure}

Before studying an N-atom chain, much understanding can be gained from a two-atom system. We consider the initial state $\ket{10}=(\ket{+}+\ket{-})/\sqrt{2}$, with $\ket{\pm}=(\ket{10}\pm\ket{01})/\sqrt{2}$. The ``dark" state $\ket{-}$ is decoupled from other states, while $\ket{+}$ is coupled to $\ket{11}$ by Rabi frequency $\sqrt{2}\Omega$ (Fig.~\ref{fig:array}(a)). Under continuous driving with the detuning matching the Rydberg-Rydberg interaction, half the population undergoes the Rabi flopping between $\ket{+}$ and $\ket{11}$, and the population of $\ket{11}$ oscillates between 0 and 0.5 at frequency $\sqrt{2}\Omega$. When the $\ket{11}$ approaches 0, the population of $\ket{10}$ and $\ket{01}$ take turns to reach the maximum, depending on the relative phase between $\ket{+}$ and $\ket{-}$. As a result, the oscillation frequency of $\ket{10}$ is half of that of $\ket{11}$, i.e. $\sqrt{2}\Omega/2\equiv \tilde{\Omega}$, which we define as an effective Rabi frequency. The Rydberg excitation hops to the other site every $T=\pi/\tilde{\Omega}$. 

Extending this picture to a 1D atomic array of $N$ atoms, the Rydberg excitation can hop either leftward or rightward. To control the directionality, we introduce alternating interatomic spacings $r_1$ and $r_2$ (Fig.~\ref{fig:array}(b)). The corresponding nearest-neighbor Rydberg-Rydberg interactions are $V_{r_1}\equiv V(\bm{r}_1)$ and $V_{r_2}\equiv V(\bm{r}_2)$. The van der Waals interaction is described by $V(\bm{r})=C_6/|\bm{r}|^6$, where $C_6$ is the van der Waals interaction coefficient (along the array direction if anisotropic). The driving laser detuning alternates between $\Delta_1=V_{r_1}$ and $\Delta_2=V_{r_2}$, completing a $\pi$ pulse with duration $T$ at each detuning (Fig.~\ref{fig:array}(c)). The resulting piecewise Hamiltonian is
\begin{equation}
    H(t)=\begin{cases}
        H_1,\quad (2l)T\leq t<(2l+1)T\\
        H_2,\quad (2l+1)T\leq t<(2l+2)T
    \end{cases}
    \label{eq1}
\end{equation}
where $l=0,1,2,\cdots$ and 
\begin{align}
        H_i=&\frac{\Omega}{2}\sum_{j=1}^{N}(e^{-i\Delta_it}\sigma_j^+ +h.c.)  +V_{r_1}\sum_{j=1}^{\left\lfloor \frac{N}{2} \right\rfloor} n_{2j-1}n_{2j} 
        \nonumber\\
        &+V_{r_2}\sum_{j=1}^{\left\lfloor \frac{N-1}{2} \right\rfloor} n_{2j}n_{2j+1} +V_{r_1+r_2} \sum_{j=1}^{N-2}n_jn_{j+2}
        \label{eq: Hamiltonian2}
\end{align}
with $n_i=\ket{1}_{ii}\bra{1}$, $\sigma_i^+=\ket{1}_{ii}\bra{0}$, and $\left\lfloor\cdots\right\rfloor$ represents the floor function. Without loss of generality, we label the chain from left to right as sites 1, 2, 3, $\cdots$ and assume it begins with an interatomic distance of $r_1$, meaning the distance between sites 1 and 2 is $r_1$, while that between sites 2 and 3 is $r_2$ with $r_1 < r_2$. Under the Hamiltonian in Eq.~\ref{eq1}, transport proceeds to the right for the Rydberg excitation in Fig.~\ref{fig:array}(b).

 We neglect interactions at distances beyond next-nearest-neighbors, due to the rapid decay of van der Waals interactions with distance. In fact, the ratio of interactions between next-nearest-neighbors to those between nearest-neighbors is no more than $\left(r_2/(r_1+r_2)\right)^6\lesssim[r_2/(2r_1)]^6=V_{r_1}/(2^6V_{r_2})$. For typical values of $V_{r_1}/V_{r_2}$ on the order of a few, the ratio amounts to only a few percent.

\begin{figure}
\centering
\includegraphics[width=0.44\textwidth]{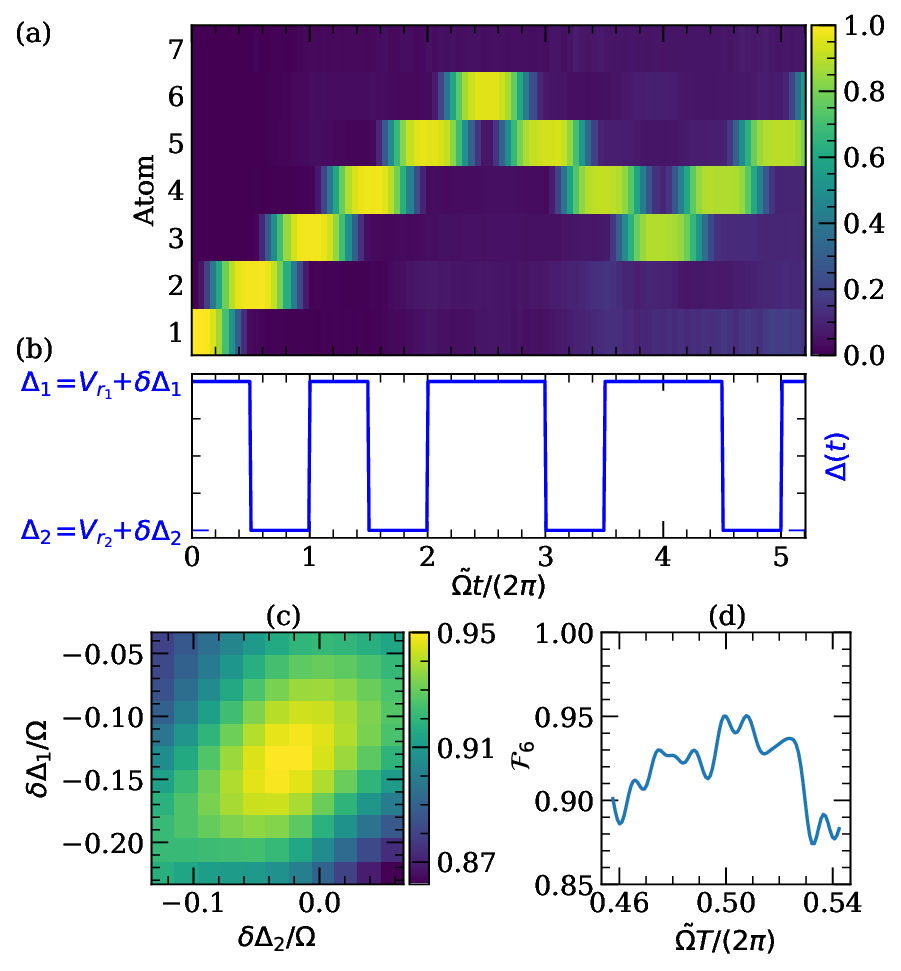}
\caption{Directional excitation transport with $V_{r_1}/\Omega=20$, $V_{r_2}/\Omega=10$, $\tilde{\Omega}T/(2\pi)=0.5$, $\delta\Delta_1/\Omega=-0.133$ and $\delta\Delta_2/\Omega=-0.033$. The color bar indicates Rydberg state population of each site. The direction of the excitation propagation (a) is controlled by the driving field pulse sequence (b). (c) Truth-table transport fidelity $\mathcal{F}_6$ as a function of detuning mismatch $\delta\Delta_i$. We determine the optimal $\delta\Delta_i$ with $\tilde{\Omega}T/(2\pi) = 0.5$. (d) A scan of $\mathcal{F}_6$ vs. $T$ at the obtained $\delta\Delta_i$ confirms that this choice of $T$ maximizes performance.}
\label{fig:propagation}
\end{figure}

\textit{Directional transport.}--Having established the theoretical framework, we now examine how excitation transport unfolds in a 1D chain through numerical simulations. 
Our simulation begins with an initial Rydberg excitation at the end of a 1D chain of 7 atoms (Fig.~\ref{fig:propagation}) and evolves under the Schr\"{o}dinger equation~\cite{johansson2012qutip,lambert2024qutip}.
Under periodic modulation of the driving laser detuning, the excitation propagates unidirectionally. However, our transport scheme is fully programmable: applying two consecutive $\pi$ pulses at the same detuning reverses the propagation direction (Fig.~\ref{fig:propagation}(a)(b)). More generally, this approach extends to routing in 2D networks, provided that energy separations are sufficiently large.

To achieve optimal transport fidelity, energy hierarchy $\Omega\ll |V_{r_2}|,  |V_{r_1}-V_{r_2}|$ must be maintained, ensuring single excitations and excitations towards the undesired direction are suppressed. The next-nearest-neighbor interaction $V_{r_1+r_2}$ is important in an equally-spaced array to suppress three consecutive excitations, i.e. $\ket{\cdots 0 1110\cdots}$. However, in our unequally-spaced array, the third atom is out of resonance even without $V_{r_1+r_2}$. Consequently, simulations show no noticeable difference with and without the $V_{r_1+r_2}$ term in the Hamiltonian (Eq.~\ref{eq: Hamiltonian2}). Our scheme therefore achieves higher transport fidelity than methods relying on kinetic constraints from $V_{r_1+r_2}$, as the finite range of van der Waals interactions makes the condition $\Omega\ll |V_{r_1+r_2}|$ challenging to satisfy. 

By analyzing the two-atom picture, we find that the best fidelity occurs at a small detuning mismatch $\delta\Delta$ in addition to the interaction strength $V$. This is because the $\ket{+}$ state is coupled not only to $\ket{11}$ but also off-resonantly to $\ket{00}$ with a detuning $\Delta$, introducing a slowly-varying envelope with frequency $\tilde\Omega^2/\Delta$. Applying a small detuning mismatch $\delta\Delta$ helps mitigate the impact of this coupling. In an array, the detuning mismatch depends not only on the driving and interaction strengths, but also on the array size, the path taken and the initial state. One factor contributing to this complexity is the small erroneous Rydberg population that appears on undesired sites. This population varies across different conditions and, in turn, affects the population on desired sites. As a result, the detuning mismatch does not have simple analytical solutions. We allow $\delta\Delta_i$ to take different values for the two interaction strengths and numerically optimize them to maximize the truth-table transport fidelity, achieving $\mathcal{F}_6=0.950$ (Fig.~\ref{fig:propagation}(c)(d)). In this definition, the input state is measured in the 1st atom basis, and after five $\pi$ pulses, the output state is measured in the 6th atom basis. With the same optimal parameters, the Rydberg population of the 6th site is also maximized ($P_6=0.956$). As interaction strengths ($|V_{r_i}|/\Omega$) increase, the transport fidelity approaches 1.
 
\begin{figure}
\centering
\includegraphics[width=0.47\textwidth]{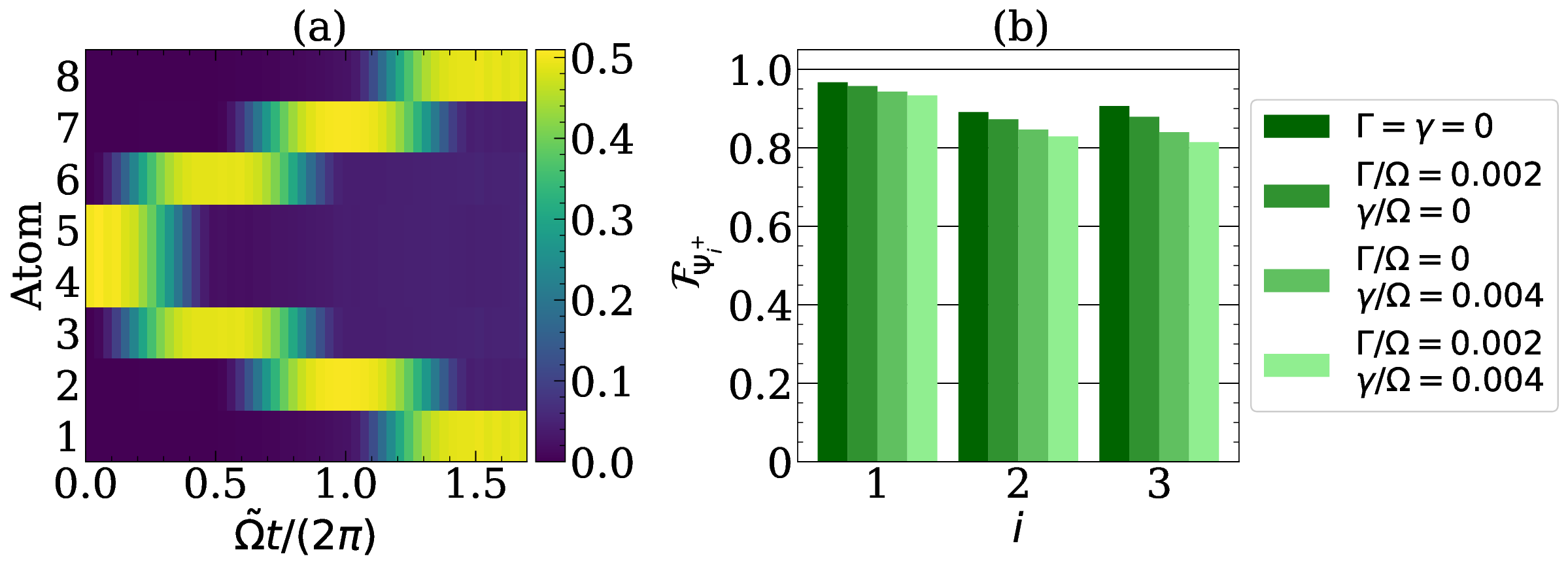}
\caption{Entanglement transfer. $\Psi_0^+$ is initiated between atom 4 and 5 and three sequential $\pi$ pulses are applied. The parameters, except $\Gamma$ and $\gamma$, are the same as in Fig.~\ref{fig:propagation}. (a) Rydberg population with $\Gamma/\Omega=0.002$ and $\gamma/\Omega=0.004$. The population of atom 1 and atom 8 after three $\pi$ pulses is 0.491. (b) Bell state fidelity $\mathcal{F}_{\Psi_i^+}$ after the $i$th ($i=1,2,3$) $\pi$ pulse with neither decay nor dephasing, only decay, only dephasing, and both decay and dephasing.}
\label{fig:Bell state}
\end{figure}

\textit{Entanglement transport.}--Next, we consider the transport of a Bell pair, $\frac{1}{\sqrt{2}}(\ket{00010000}+\ket{00001000})$. The same pulse sequence used to transport an excitation from site 4 to site 1 can also transfer an excitation from site 5 to site 8 (Fig.~\ref{fig:Bell state}(a)). This enables relocation of entanglement, initially generated between neighboring sites, to sites at opposite ends of the chain. The population transfer exhibits similar fidelity to that of a single excitation hopping the same number of steps.  
However, in entanglement transfer, population transport fidelity alone is not the key figure of merit; coherence is central. To assess this, we incorporate realistic decay and dephasing mechanisms and simulate the dynamics using the Lindblad master equation:
\begin{align}
    &\frac{d\rho}{dt}=i[\rho,H]   +\mathcal{L}_{decay}(\rho) +\mathcal{L}_{deph}(\rho) 
\end{align}
where $\rho$ is the density matrix of the atomic chain, and $\mathcal{L}_{decay}(\rho)$ and $\mathcal{L}_{deph}(\rho)$ are the Lindblad superoperators describing decay and dephasing, respectively. Assuming each Rydberg atom decay with rate $\Gamma$, and the ground-Rydberg atom dephasing rate $\gamma$,
\begin{subequations}
\begin{align}
\mathcal{L}_{decay}(\rho) &= \frac{\Gamma}{2}\sum_{k=1}^{N}  \left( 2 \sigma^-_k \rho \sigma^+_k - \sigma^+_k \sigma^-_k \rho - \rho \sigma^+_k \sigma^-_k  \right)
\\
\mathcal{L}_{deph}(\rho) &= \frac{\gamma}{2}\sum_{k=1}^{N}  \left( \sigma^z_k \rho \sigma^z_k - \rho    \right)
\end{align}
\end{subequations}

We evaluate entanglement transfer using the Bell state fidelity $\mathcal{F}_{\Psi_i^+}=\bra{\Psi_i^+}\rho_i\ket{\Psi^+_i}$ (Fig.~\ref{fig:Bell state}(b)), where $\rho_i$ is the reduced two-atom density matrix of atom $4-i$ and atom $5+i$ and $\ket{\Psi^+_i}\equiv(\ket{0}_{4-i}\ket{1}_{5+i}+\ket{1}_{4-i}\ket{0}_{5+i})/\sqrt{2}$ is the target Bell state after the $i$th $\pi$ pulse.
To evaluate the feasibility of our scheme under realistic conditions, we consider implementing the scheme using $^{87}$Rb tweezer arrays.
The Rydberg excitation can be realized through a one-photon excitation to a P state, or more commonly through a two-photon process via a low-lying intermediate state 5P or 6P. The detuning from the intermediate state must be sufficiently large such that the increased decay rate due to the admixture of the intermediate state does not derail the transport fidelity. The choice of $\Gamma/\Omega=0.2\%$ in simulations is motivated by an excitation scheme detuned from 6P by roughly 1~GHz, in combination with $\Omega/(2\pi)=3$~MHz. We also account for dephasing effects, represented by $\gamma$, which includes finite laser linewidth and stray electric field noise. While these sources of decoherence have minimal impact on Rydberg population transfer, they introduce a noticeable reduction in Bell state fidelity. Despite this, our approach remains highly effective for entanglement transport. After three hopping steps, taking $0.7~\mu$s, the distance between the two entangled atoms is roughly 80~$\mu$m, for interatomic distances $r_i\gtrsim10~\mu$m. The average transport speed is two orders of magnitude faster than mechanical atom transport~\cite{bluvstein2022quantum}, and the protocol’s efficiency for short-distance quantum information transfer may surpass that of photonic links.


\begin{figure}[ht]
\centering
\includegraphics[width=0.45\textwidth]{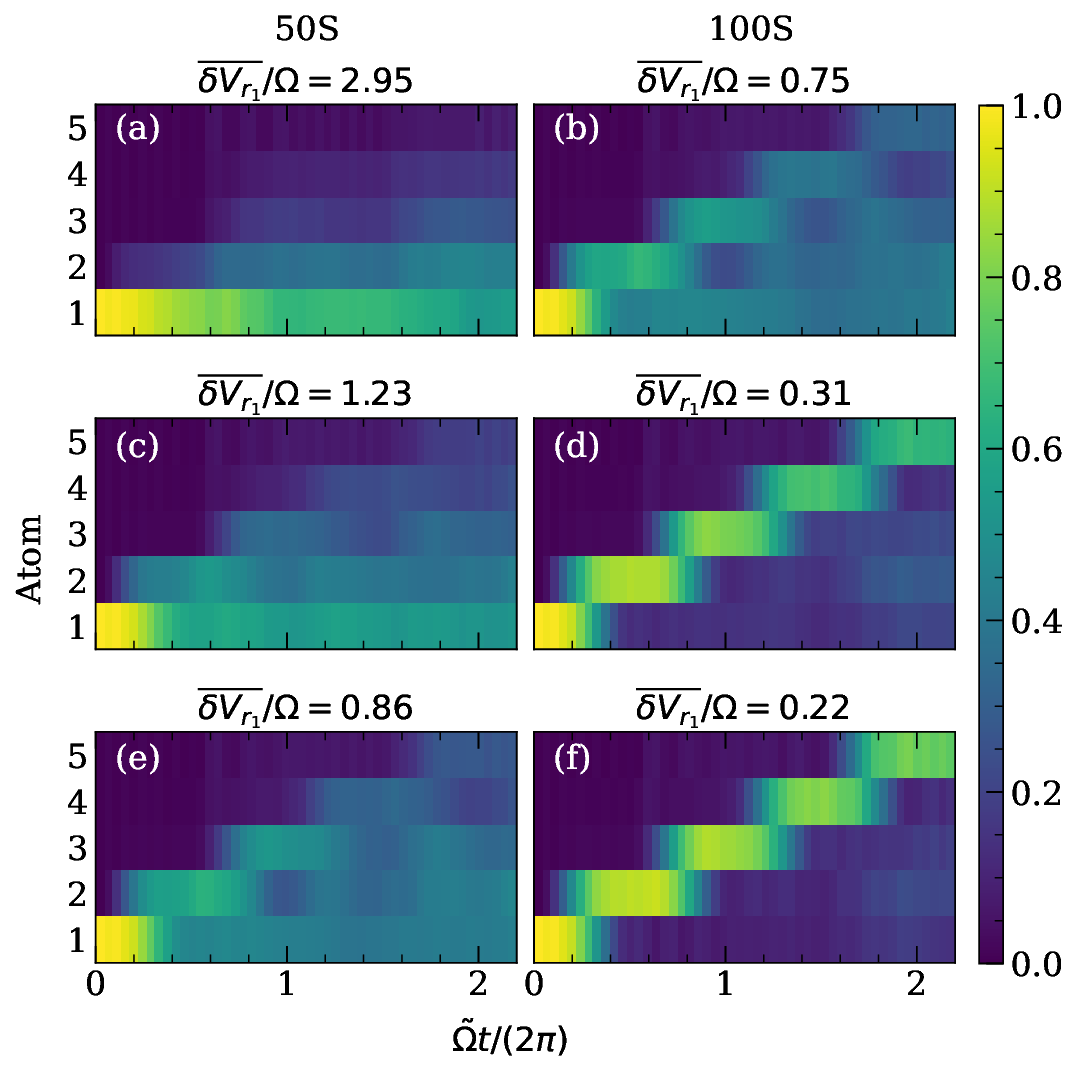}
\caption{Impact of position disorder on transport fidelity. The parameters are $V_{r_1}/\Omega=8.4$, $V_{r_2}/\Omega=4.2$, $\Gamma/\Omega=0.002$, $\gamma/\Omega=0.004$, $\tilde{\Omega}T/(2\pi)=0.5$, $\delta\Delta_1/\Omega=-0.293$ and $\delta\Delta_2/\Omega=-0.267$. We average over 50 realizations, where each site's position deviation from its ideal position is randomly sampled from a Gaussian distributions with standard deviation $\bm{\sigma}=(\sigma_x,\sigma_y,\sigma_z)$. The values are: $\bm{\sigma}=(120,120,1080)~$nm for (a,b), (50,50,450)~nm for (c,d) and (35,35,315)~nm for (e,f).}
\label{fig:disorder}
\end{figure}

\textit{Position disorder.}--A key challenge in realizing this scheme is position disorder~\cite{zhao2024observation,marcuzzi2017facilitation,valencia2024rydberg}. A common experimental practice in Rydberg experiments is to switch off atomic tweezer traps before the dynamics begin.
At sufficiently low temperature $T_{temp}$ (a few tens of microkelvins), atomic positions remain approximately constant during a single experimental realization (on the order of a microsecond). Specifically, we randomly sample the displacements $\delta\bm{r}_j$ from the ideal atomic positions according to a Gaussian distribution with zero mean and standard deviation $\sigma_l=\sqrt{\frac{k_BT_{temp}}{m\omega_l^2}}$ along the $l$ axis, where $\omega_l$ is the trapping frequency and $m$ is the atomic mass. 
Fig.~\ref{fig:disorder} compares the performances of 50S and 100S states. For a typical 852~nm optical tweezer with a $1/e^2$ waist of 1.7~$\mu$m and trap depth of 2.6~mK, the ratio of the axial to radial position spread is $\sigma_z/\sigma_x=9$. Standard optical molasses typically cool atoms to temperatures around $50~\mu$K, resulting in a position spread $\sigma_x=120~$nm (Fig.~\ref{fig:disorder}(a,b)) given the above trap parameters. Recently, a simple and efficient cooling method $\Lambda-$enhanced grey molasses~\cite{evered2023high,rosi2018lambda} cools atoms to $10~\mu$K (Fig.~\ref{fig:disorder}(c,d)). With Raman sideband cooling~\cite{10.21468/SciPostPhys.10.3.052} approaching the motional ground state, the position spread is estimated to be around $\sigma_x=35~$nm (Fig.~\ref{fig:disorder}(e,f)). To make a compromise between position disorder and the energy hierarchy $\Omega\ll |V_{r_2}|,  |V_{r_1}-V_{r_2}|$, we reduce the interaction strengths $|V_{r_i}|/\Omega$ by a factor of $\sim$2. With these interaction strengths, the transport fidelity is comparably affected by both position disorder and weaker separation of energy scales.

The position disorder leads to deviation in the interaction energy, approximated as 
\begin{equation}
    \overline{\delta V_{r_i}}\approx6 |V_{r_i}|^{7/6}\frac{\sqrt{2}\sigma_x}{|C_6|^{1/6}},
    \label{eq: interaction disorder}
\end{equation}
where $x$ denotes the array direction. This shows that for given interaction strengths ($|V_{r_i}|$) and position disorder, larger van der Waals interaction coefficient ($|C_6|$) mitigates energy deviations. This motivates using high principal quantum number $n$, which reduces sensitivity to position disorder while also increasing the hopping distance per step (larger interatomic distance for given interaction strength) and extending Rydberg lifetimes. However, high $n$ increases sensitivity to stray electric fields~\cite{panja2024electric,Ananddualspeciesrydberg,wilson2022trapping}, with reasonable choices typically in the range $n\sim70-100$.

Given the challenges posed by position disorder, high $n$ and advanced cooling are essential to avoid localization. Since $C_6$ scales as $n^{11}$, choosing $n=100$ improves the figure of merit $\overline{\delta V_{r_1}}/\Omega$ (Eq.~\ref{eq: interaction disorder}) by approximately a factor of 4 compared to $n=50$. Ballistic transport, requiring $\overline{\delta V_{r_1}}/\Omega\lesssim0.3$, is achievable with $\Lambda-$enhanced grey molasses. The ratio $\overline{\delta V_{r_1}}/\Omega$ is largely insensitive to $\Omega$ with fixed $|V_{r_i}|/\Omega$, meaning that slowing down the entire dynamics does not significantly suppress interaction energy disorder.
For the 100S state, driven at Rabi frequency $\Omega/(2\pi)=3$~MHz, the interatomic distances of the atomic array are $r_1=11.4~\mu$m and $r_2=12.8~\mu$m. The resulting average transport speed (51~$\mu$m/$\mu$s) is consistent with earlier discusions demonstrating that a few hopping steps suffice to move atoms over distances relevant for zoned quantum information processing tasks~\cite{bluvstein2024logical}.



\textit{Additional experimental considerations.}--To prepare the initial state for the transport process, we can use either local addressing~\cite{graham2022multi,omran2019generation} or tweezer repositioning~\footnote{ The array can be repositioned to locations that satisfy facilitation conditions required by our transport protocol after the completion of a global excitation pulse.}
to create a Rydberg excitation. With the help of Rydberg blockade, this approach also enables the preparation of a Bell pair~\cite{madjarov2020high,PhysRevLett.123.170503,thompsongate}.
The alternating detuning can be realized by using an RF switch to send alternating RF tones to an acoustic optic modulator. The typical switching time scale is around 20~ns (8.5\% of the $\pi$-pulse duration for our chosen parameters). In practice, the pulse intensity may be switched off during the switching time, or more sophisticated pulse shaping may be employed to mitigate experimental imperfections. Additionally, each step of the transport process can be optimized individually, rather than sharing identical parameters across all steps.



\textit{Summary and outlook.}--In conclusion, we have proposed an experimentally feasible scheme for programmable directional transport of Rydberg excitations and entangled states using Rydberg antiblockade and modulated pulses. While disorder typically localizes excitations, our numerical simulations show that coherent transport remains robust against small position disorders, achievable with state-of-the-art experimental setups. This opens up new possibilities for investigating the interplay of dissipation and disorder in nonequilibrium quantum dynamics. Future investigations could explore the impact of atom trapping on the dynamics and extending the protocol to 2D networks with multiple excitations, potentially offering new avenues for quantum information processing.


\textit{Acknowledgments.}--We thank Chen-Lung Hung and Qi Zhou for insightful discussions. This work was supported by Purdue startup fund and AFOSR Grant FA9550-22-1-0327.


\bibliographystyle{apsrev4-2}
\bibliography{references}

\begin{thebibliography}{45}%
\makeatletter
\providecommand \@ifxundefined [1]{%
 \@ifx{#1\undefined}
}%
\providecommand \@ifnum [1]{%
 \ifnum #1\expandafter \@firstoftwo
 \else \expandafter \@secondoftwo
 \fi
}%
\providecommand \@ifx [1]{%
 \ifx #1\expandafter \@firstoftwo
 \else \expandafter \@secondoftwo
 \fi
}%
\providecommand \natexlab [1]{#1}%
\providecommand \enquote  [1]{``#1''}%
\providecommand \bibnamefont  [1]{#1}%
\providecommand \bibfnamefont [1]{#1}%
\providecommand \citenamefont [1]{#1}%
\providecommand \href@noop [0]{\@secondoftwo}%
\providecommand \href [0]{\begingroup \@sanitize@url \@href}%
\providecommand \@href[1]{\@@startlink{#1}\@@href}%
\providecommand \@@href[1]{\endgroup#1\@@endlink}%
\providecommand \@sanitize@url [0]{\catcode `\\12\catcode `\$12\catcode `\&12\catcode `\#12\catcode `\^12\catcode `\_12\catcode `\%12\relax}%
\providecommand \@@startlink[1]{}%
\providecommand \@@endlink[0]{}%
\providecommand \url  [0]{\begingroup\@sanitize@url \@url }%
\providecommand \@url [1]{\endgroup\@href {#1}{\urlprefix }}%
\providecommand \urlprefix  [0]{URL }%
\providecommand \Eprint [0]{\href }%
\providecommand \doibase [0]{https://doi.org/}%
\providecommand \selectlanguage [0]{\@gobble}%
\providecommand \bibinfo  [0]{\@secondoftwo}%
\providecommand \bibfield  [0]{\@secondoftwo}%
\providecommand \translation [1]{[#1]}%
\providecommand \BibitemOpen [0]{}%
\providecommand \bibitemStop [0]{}%
\providecommand \bibitemNoStop [0]{.\EOS\space}%
\providecommand \EOS [0]{\spacefactor3000\relax}%
\providecommand \BibitemShut  [1]{\csname bibitem#1\endcsname}%
\let\auto@bib@innerbib\@empty
\bibitem [{\citenamefont {Claassen}\ \emph {et~al.}(2017)\citenamefont {Claassen}, \citenamefont {Jiang}, \citenamefont {Moritz},\ and\ \citenamefont {Devereaux}}]{claassen2017dynamical}%
  \BibitemOpen
  \bibfield  {author} {\bibinfo {author} {\bibfnamefont {M.}~\bibnamefont {Claassen}}, \bibinfo {author} {\bibfnamefont {H.-C.}\ \bibnamefont {Jiang}}, \bibinfo {author} {\bibfnamefont {B.}~\bibnamefont {Moritz}},\ and\ \bibinfo {author} {\bibfnamefont {T.~P.}\ \bibnamefont {Devereaux}},\ }\href@noop {} {\bibfield  {journal} {\bibinfo  {journal} {Nature Communications}\ }\textbf {\bibinfo {volume} {8}},\ \bibinfo {pages} {1192} (\bibinfo {year} {2017})}\BibitemShut {NoStop}%
\bibitem [{\citenamefont {Sun}\ \emph {et~al.}(2023)\citenamefont {Sun}, \citenamefont {Goldman}, \citenamefont {Aidelsburger},\ and\ \citenamefont {Bukov}}]{sun2023engineering}%
  \BibitemOpen
  \bibfield  {author} {\bibinfo {author} {\bibfnamefont {B.-Y.}\ \bibnamefont {Sun}}, \bibinfo {author} {\bibfnamefont {N.}~\bibnamefont {Goldman}}, \bibinfo {author} {\bibfnamefont {M.}~\bibnamefont {Aidelsburger}},\ and\ \bibinfo {author} {\bibfnamefont {M.}~\bibnamefont {Bukov}},\ }\href@noop {} {\bibfield  {journal} {\bibinfo  {journal} {PRX Quantum}\ }\textbf {\bibinfo {volume} {4}},\ \bibinfo {pages} {020329} (\bibinfo {year} {2023})}\BibitemShut {NoStop}%
\bibitem [{\citenamefont {Wintersperger}\ \emph {et~al.}(2020)\citenamefont {Wintersperger}, \citenamefont {Braun}, \citenamefont {{\"U}nal}, \citenamefont {Eckardt}, \citenamefont {Liberto}, \citenamefont {Goldman}, \citenamefont {Bloch},\ and\ \citenamefont {Aidelsburger}}]{wintersperger2020realization}%
  \BibitemOpen
  \bibfield  {author} {\bibinfo {author} {\bibfnamefont {K.}~\bibnamefont {Wintersperger}}, \bibinfo {author} {\bibfnamefont {C.}~\bibnamefont {Braun}}, \bibinfo {author} {\bibfnamefont {F.~N.}\ \bibnamefont {{\"U}nal}}, \bibinfo {author} {\bibfnamefont {A.}~\bibnamefont {Eckardt}}, \bibinfo {author} {\bibfnamefont {M.~D.}\ \bibnamefont {Liberto}}, \bibinfo {author} {\bibfnamefont {N.}~\bibnamefont {Goldman}}, \bibinfo {author} {\bibfnamefont {I.}~\bibnamefont {Bloch}},\ and\ \bibinfo {author} {\bibfnamefont {M.}~\bibnamefont {Aidelsburger}},\ }\href@noop {} {\bibfield  {journal} {\bibinfo  {journal} {Nature Physics}\ }\textbf {\bibinfo {volume} {16}},\ \bibinfo {pages} {1058} (\bibinfo {year} {2020})}\BibitemShut {NoStop}%
\bibitem [{\citenamefont {Lodahl}\ \emph {et~al.}(2017)\citenamefont {Lodahl}, \citenamefont {Mahmoodian}, \citenamefont {Stobbe}, \citenamefont {Rauschenbeutel}, \citenamefont {Schneeweiss}, \citenamefont {Volz}, \citenamefont {Pichler},\ and\ \citenamefont {Zoller}}]{lodahl2017chiral}%
  \BibitemOpen
  \bibfield  {author} {\bibinfo {author} {\bibfnamefont {P.}~\bibnamefont {Lodahl}}, \bibinfo {author} {\bibfnamefont {S.}~\bibnamefont {Mahmoodian}}, \bibinfo {author} {\bibfnamefont {S.}~\bibnamefont {Stobbe}}, \bibinfo {author} {\bibfnamefont {A.}~\bibnamefont {Rauschenbeutel}}, \bibinfo {author} {\bibfnamefont {P.}~\bibnamefont {Schneeweiss}}, \bibinfo {author} {\bibfnamefont {J.}~\bibnamefont {Volz}}, \bibinfo {author} {\bibfnamefont {H.}~\bibnamefont {Pichler}},\ and\ \bibinfo {author} {\bibfnamefont {P.}~\bibnamefont {Zoller}},\ }\href@noop {} {\bibfield  {journal} {\bibinfo  {journal} {Nature}\ }\textbf {\bibinfo {volume} {541}},\ \bibinfo {pages} {473} (\bibinfo {year} {2017})}\BibitemShut {NoStop}%
\bibitem [{\citenamefont {McDonald}\ \emph {et~al.}(2018)\citenamefont {McDonald}, \citenamefont {Pereg-Barnea},\ and\ \citenamefont {Clerk}}]{mcdonald2018phase}%
  \BibitemOpen
  \bibfield  {author} {\bibinfo {author} {\bibfnamefont {A.}~\bibnamefont {McDonald}}, \bibinfo {author} {\bibfnamefont {T.}~\bibnamefont {Pereg-Barnea}},\ and\ \bibinfo {author} {\bibfnamefont {A.}~\bibnamefont {Clerk}},\ }\href@noop {} {\bibfield  {journal} {\bibinfo  {journal} {Physical Review X}\ }\textbf {\bibinfo {volume} {8}},\ \bibinfo {pages} {041031} (\bibinfo {year} {2018})}\BibitemShut {NoStop}%
\bibitem [{\citenamefont {Guo}\ \emph {et~al.}(2022)\citenamefont {Guo}, \citenamefont {Putzke}, \citenamefont {Konyzheva}, \citenamefont {Huang}, \citenamefont {Gutierrez-Amigo}, \citenamefont {Errea}, \citenamefont {Chen}, \citenamefont {Vergniory}, \citenamefont {Felser}, \citenamefont {Fischer} \emph {et~al.}}]{guo2022switchable}%
  \BibitemOpen
  \bibfield  {author} {\bibinfo {author} {\bibfnamefont {C.}~\bibnamefont {Guo}}, \bibinfo {author} {\bibfnamefont {C.}~\bibnamefont {Putzke}}, \bibinfo {author} {\bibfnamefont {S.}~\bibnamefont {Konyzheva}}, \bibinfo {author} {\bibfnamefont {X.}~\bibnamefont {Huang}}, \bibinfo {author} {\bibfnamefont {M.}~\bibnamefont {Gutierrez-Amigo}}, \bibinfo {author} {\bibfnamefont {I.}~\bibnamefont {Errea}}, \bibinfo {author} {\bibfnamefont {D.}~\bibnamefont {Chen}}, \bibinfo {author} {\bibfnamefont {M.~G.}\ \bibnamefont {Vergniory}}, \bibinfo {author} {\bibfnamefont {C.}~\bibnamefont {Felser}}, \bibinfo {author} {\bibfnamefont {M.~H.}\ \bibnamefont {Fischer}}, \emph {et~al.},\ }\href@noop {} {\bibfield  {journal} {\bibinfo  {journal} {Nature}\ }\textbf {\bibinfo {volume} {611}},\ \bibinfo {pages} {461} (\bibinfo {year} {2022})}\BibitemShut {NoStop}%
\bibitem [{\citenamefont {Li}\ \emph {et~al.}(2022)\citenamefont {Li}, \citenamefont {You}, \citenamefont {Shao},\ and\ \citenamefont {Li}}]{li2022coherent}%
  \BibitemOpen
  \bibfield  {author} {\bibinfo {author} {\bibfnamefont {X.}~\bibnamefont {Li}}, \bibinfo {author} {\bibfnamefont {J.}~\bibnamefont {You}}, \bibinfo {author} {\bibfnamefont {X.}~\bibnamefont {Shao}},\ and\ \bibinfo {author} {\bibfnamefont {W.}~\bibnamefont {Li}},\ }\href@noop {} {\bibfield  {journal} {\bibinfo  {journal} {Physical Review A}\ }\textbf {\bibinfo {volume} {105}},\ \bibinfo {pages} {032417} (\bibinfo {year} {2022})}\BibitemShut {NoStop}%
\bibitem [{\citenamefont {Banuls}\ \emph {et~al.}(2020)\citenamefont {Banuls}, \citenamefont {Blatt}, \citenamefont {Catani}, \citenamefont {Celi}, \citenamefont {Cirac}, \citenamefont {Dalmonte}, \citenamefont {Fallani}, \citenamefont {Jansen}, \citenamefont {Lewenstein}, \citenamefont {Montangero} \emph {et~al.}}]{banuls2020simulating}%
  \BibitemOpen
  \bibfield  {author} {\bibinfo {author} {\bibfnamefont {M.~C.}\ \bibnamefont {Banuls}}, \bibinfo {author} {\bibfnamefont {R.}~\bibnamefont {Blatt}}, \bibinfo {author} {\bibfnamefont {J.}~\bibnamefont {Catani}}, \bibinfo {author} {\bibfnamefont {A.}~\bibnamefont {Celi}}, \bibinfo {author} {\bibfnamefont {J.~I.}\ \bibnamefont {Cirac}}, \bibinfo {author} {\bibfnamefont {M.}~\bibnamefont {Dalmonte}}, \bibinfo {author} {\bibfnamefont {L.}~\bibnamefont {Fallani}}, \bibinfo {author} {\bibfnamefont {K.}~\bibnamefont {Jansen}}, \bibinfo {author} {\bibfnamefont {M.}~\bibnamefont {Lewenstein}}, \bibinfo {author} {\bibfnamefont {S.}~\bibnamefont {Montangero}}, \emph {et~al.},\ }\href@noop {} {\bibfield  {journal} {\bibinfo  {journal} {The European Physical Journal D}\ }\textbf {\bibinfo {volume} {74}},\ \bibinfo {pages} {1} (\bibinfo {year} {2020})}\BibitemShut {NoStop}%
\bibitem [{\citenamefont {Almanakly}\ \emph {et~al.}(2024)\citenamefont {Almanakly}, \citenamefont {Yankelevich}, \citenamefont {Hays}, \citenamefont {Kannan}, \citenamefont {Assouly}, \citenamefont {Greene}, \citenamefont {Gingras}, \citenamefont {Niedzielski}, \citenamefont {Stickler}, \citenamefont {Schwartz} \emph {et~al.}}]{almanakly2024deterministic}%
  \BibitemOpen
  \bibfield  {author} {\bibinfo {author} {\bibfnamefont {A.}~\bibnamefont {Almanakly}}, \bibinfo {author} {\bibfnamefont {B.}~\bibnamefont {Yankelevich}}, \bibinfo {author} {\bibfnamefont {M.}~\bibnamefont {Hays}}, \bibinfo {author} {\bibfnamefont {B.}~\bibnamefont {Kannan}}, \bibinfo {author} {\bibfnamefont {R.}~\bibnamefont {Assouly}}, \bibinfo {author} {\bibfnamefont {A.}~\bibnamefont {Greene}}, \bibinfo {author} {\bibfnamefont {M.}~\bibnamefont {Gingras}}, \bibinfo {author} {\bibfnamefont {B.~M.}\ \bibnamefont {Niedzielski}}, \bibinfo {author} {\bibfnamefont {H.}~\bibnamefont {Stickler}}, \bibinfo {author} {\bibfnamefont {M.~E.}\ \bibnamefont {Schwartz}}, \emph {et~al.},\ }\href@noop {} {\bibfield  {journal} {\bibinfo  {journal} {arXiv preprint arXiv:2408.05164}\ } (\bibinfo {year} {2024})}\BibitemShut {NoStop}%
\bibitem [{\citenamefont {Grankin}\ \emph {et~al.}(2018)\citenamefont {Grankin}, \citenamefont {Guimond}, \citenamefont {Vasilyev}, \citenamefont {Vermersch},\ and\ \citenamefont {Zoller}}]{grankin2018free}%
  \BibitemOpen
  \bibfield  {author} {\bibinfo {author} {\bibfnamefont {A.}~\bibnamefont {Grankin}}, \bibinfo {author} {\bibfnamefont {P.}~\bibnamefont {Guimond}}, \bibinfo {author} {\bibfnamefont {D.}~\bibnamefont {Vasilyev}}, \bibinfo {author} {\bibfnamefont {B.}~\bibnamefont {Vermersch}},\ and\ \bibinfo {author} {\bibfnamefont {P.}~\bibnamefont {Zoller}},\ }\href@noop {} {\bibfield  {journal} {\bibinfo  {journal} {Physical Review A}\ }\textbf {\bibinfo {volume} {98}},\ \bibinfo {pages} {043825} (\bibinfo {year} {2018})}\BibitemShut {NoStop}%
\bibitem [{\citenamefont {Wang}\ \emph {et~al.}(2024)\citenamefont {Wang}, \citenamefont {Zhang}, \citenamefont {Jiao}, \citenamefont {Zhang}, \citenamefont {Lu}, \citenamefont {Li}, \citenamefont {Qiu},\ and\ \citenamefont {Jing}}]{wang2024quantum}%
  \BibitemOpen
  \bibfield  {author} {\bibinfo {author} {\bibfnamefont {J.}~\bibnamefont {Wang}}, \bibinfo {author} {\bibfnamefont {Q.}~\bibnamefont {Zhang}}, \bibinfo {author} {\bibfnamefont {Y.-F.}\ \bibnamefont {Jiao}}, \bibinfo {author} {\bibfnamefont {S.-D.}\ \bibnamefont {Zhang}}, \bibinfo {author} {\bibfnamefont {T.-X.}\ \bibnamefont {Lu}}, \bibinfo {author} {\bibfnamefont {Z.}~\bibnamefont {Li}}, \bibinfo {author} {\bibfnamefont {C.-W.}\ \bibnamefont {Qiu}},\ and\ \bibinfo {author} {\bibfnamefont {H.}~\bibnamefont {Jing}},\ }\href@noop {} {\bibfield  {journal} {\bibinfo  {journal} {arXiv preprint arXiv:2403.09979}\ } (\bibinfo {year} {2024})}\BibitemShut {NoStop}%
\bibitem [{\citenamefont {Xie}\ and\ \citenamefont {Xu}(2024)}]{xie2024quantum}%
  \BibitemOpen
  \bibfield  {author} {\bibinfo {author} {\bibfnamefont {D.}~\bibnamefont {Xie}}\ and\ \bibinfo {author} {\bibfnamefont {C.}~\bibnamefont {Xu}},\ }\href@noop {} {\bibfield  {journal} {\bibinfo  {journal} {Physical Review Applied}\ }\textbf {\bibinfo {volume} {22}},\ \bibinfo {pages} {064072} (\bibinfo {year} {2024})}\BibitemShut {NoStop}%
\bibitem [{\citenamefont {Valencia-Tortora}\ \emph {et~al.}(2024)\citenamefont {Valencia-Tortora}, \citenamefont {Pancotti}, \citenamefont {Fleischhauer}, \citenamefont {Bernien},\ and\ \citenamefont {Marino}}]{valencia2024rydberg}%
  \BibitemOpen
  \bibfield  {author} {\bibinfo {author} {\bibfnamefont {R.~J.}\ \bibnamefont {Valencia-Tortora}}, \bibinfo {author} {\bibfnamefont {N.}~\bibnamefont {Pancotti}}, \bibinfo {author} {\bibfnamefont {M.}~\bibnamefont {Fleischhauer}}, \bibinfo {author} {\bibfnamefont {H.}~\bibnamefont {Bernien}},\ and\ \bibinfo {author} {\bibfnamefont {J.}~\bibnamefont {Marino}},\ }\href@noop {} {\bibfield  {journal} {\bibinfo  {journal} {Physical Review Letters}\ }\textbf {\bibinfo {volume} {132}},\ \bibinfo {pages} {223201} (\bibinfo {year} {2024})}\BibitemShut {NoStop}%
\bibitem [{\citenamefont {Kitson}\ \emph {et~al.}(2024)\citenamefont {Kitson}, \citenamefont {Haug}, \citenamefont {La~Magna}, \citenamefont {Morsch},\ and\ \citenamefont {Amico}}]{kitson2024rydberg}%
  \BibitemOpen
  \bibfield  {author} {\bibinfo {author} {\bibfnamefont {P.}~\bibnamefont {Kitson}}, \bibinfo {author} {\bibfnamefont {T.}~\bibnamefont {Haug}}, \bibinfo {author} {\bibfnamefont {A.}~\bibnamefont {La~Magna}}, \bibinfo {author} {\bibfnamefont {O.}~\bibnamefont {Morsch}},\ and\ \bibinfo {author} {\bibfnamefont {L.}~\bibnamefont {Amico}},\ }\href@noop {} {\bibfield  {journal} {\bibinfo  {journal} {Physical Review A}\ }\textbf {\bibinfo {volume} {110}},\ \bibinfo {pages} {043304} (\bibinfo {year} {2024})}\BibitemShut {NoStop}%
\bibitem [{\citenamefont {Ates}\ \emph {et~al.}(2007)\citenamefont {Ates}, \citenamefont {Pohl}, \citenamefont {Pattard},\ and\ \citenamefont {Rost}}]{ates2007antiblockade}%
  \BibitemOpen
  \bibfield  {author} {\bibinfo {author} {\bibfnamefont {C.}~\bibnamefont {Ates}}, \bibinfo {author} {\bibfnamefont {T.}~\bibnamefont {Pohl}}, \bibinfo {author} {\bibfnamefont {T.}~\bibnamefont {Pattard}},\ and\ \bibinfo {author} {\bibfnamefont {J.~M.}\ \bibnamefont {Rost}},\ }\href@noop {} {\bibfield  {journal} {\bibinfo  {journal} {Physical Review Letters}\ }\textbf {\bibinfo {volume} {98}},\ \bibinfo {pages} {023002} (\bibinfo {year} {2007})}\BibitemShut {NoStop}%
\bibitem [{\citenamefont {Amthor}\ \emph {et~al.}(2010)\citenamefont {Amthor}, \citenamefont {Giese}, \citenamefont {Hofmann},\ and\ \citenamefont {Weidem{\"u}ller}}]{amthor2010evidence}%
  \BibitemOpen
  \bibfield  {author} {\bibinfo {author} {\bibfnamefont {T.}~\bibnamefont {Amthor}}, \bibinfo {author} {\bibfnamefont {C.}~\bibnamefont {Giese}}, \bibinfo {author} {\bibfnamefont {C.~S.}\ \bibnamefont {Hofmann}},\ and\ \bibinfo {author} {\bibfnamefont {M.}~\bibnamefont {Weidem{\"u}ller}},\ }\href@noop {} {\bibfield  {journal} {\bibinfo  {journal} {Physical Review Letters}\ }\textbf {\bibinfo {volume} {104}},\ \bibinfo {pages} {013001} (\bibinfo {year} {2010})}\BibitemShut {NoStop}%
\bibitem [{\citenamefont {Marcuzzi}\ \emph {et~al.}(2017)\citenamefont {Marcuzzi}, \citenamefont {Min{\'a}{\v{r}}}, \citenamefont {Barredo}, \citenamefont {De~L{\'e}s{\'e}leuc}, \citenamefont {Labuhn}, \citenamefont {Lahaye}, \citenamefont {Browaeys}, \citenamefont {Levi},\ and\ \citenamefont {Lesanovsky}}]{marcuzzi2017facilitation}%
  \BibitemOpen
  \bibfield  {author} {\bibinfo {author} {\bibfnamefont {M.}~\bibnamefont {Marcuzzi}}, \bibinfo {author} {\bibfnamefont {J.}~\bibnamefont {Min{\'a}{\v{r}}}}, \bibinfo {author} {\bibfnamefont {D.}~\bibnamefont {Barredo}}, \bibinfo {author} {\bibfnamefont {S.}~\bibnamefont {De~L{\'e}s{\'e}leuc}}, \bibinfo {author} {\bibfnamefont {H.}~\bibnamefont {Labuhn}}, \bibinfo {author} {\bibfnamefont {T.}~\bibnamefont {Lahaye}}, \bibinfo {author} {\bibfnamefont {A.}~\bibnamefont {Browaeys}}, \bibinfo {author} {\bibfnamefont {E.}~\bibnamefont {Levi}},\ and\ \bibinfo {author} {\bibfnamefont {I.}~\bibnamefont {Lesanovsky}},\ }\href@noop {} {\bibfield  {journal} {\bibinfo  {journal} {Physical Review Letters}\ }\textbf {\bibinfo {volume} {118}},\ \bibinfo {pages} {063606} (\bibinfo {year} {2017})}\BibitemShut {NoStop}%
\bibitem [{\citenamefont {Zhao}\ \emph {et~al.}(2024)\citenamefont {Zhao}, \citenamefont {Datla}, \citenamefont {Tian}, \citenamefont {Aliyu},\ and\ \citenamefont {Loh}}]{zhao2024observation}%
  \BibitemOpen
  \bibfield  {author} {\bibinfo {author} {\bibfnamefont {L.}~\bibnamefont {Zhao}}, \bibinfo {author} {\bibfnamefont {P.~R.}\ \bibnamefont {Datla}}, \bibinfo {author} {\bibfnamefont {W.}~\bibnamefont {Tian}}, \bibinfo {author} {\bibfnamefont {M.~M.}\ \bibnamefont {Aliyu}},\ and\ \bibinfo {author} {\bibfnamefont {H.}~\bibnamefont {Loh}},\ }\href@noop {} {\bibfield  {journal} {\bibinfo  {journal} {arXiv preprint arXiv:2403.09517}\ } (\bibinfo {year} {2024})}\BibitemShut {NoStop}%
\bibitem [{\citenamefont {Liu}\ \emph {et~al.}(2022)\citenamefont {Liu}, \citenamefont {Yang}, \citenamefont {Bienias}, \citenamefont {Iadecola},\ and\ \citenamefont {Gorshkov}}]{liu2022localization}%
  \BibitemOpen
  \bibfield  {author} {\bibinfo {author} {\bibfnamefont {F.}~\bibnamefont {Liu}}, \bibinfo {author} {\bibfnamefont {Z.-C.}\ \bibnamefont {Yang}}, \bibinfo {author} {\bibfnamefont {P.}~\bibnamefont {Bienias}}, \bibinfo {author} {\bibfnamefont {T.}~\bibnamefont {Iadecola}},\ and\ \bibinfo {author} {\bibfnamefont {A.~V.}\ \bibnamefont {Gorshkov}},\ }\href@noop {} {\bibfield  {journal} {\bibinfo  {journal} {Physical Review Letters}\ }\textbf {\bibinfo {volume} {128}},\ \bibinfo {pages} {013603} (\bibinfo {year} {2022})}\BibitemShut {NoStop}%
\bibitem [{\citenamefont {Nill}\ \emph {et~al.}(2024)\citenamefont {Nill}, \citenamefont {Cabot}, \citenamefont {Trautmann}, \citenamefont {Gro{\ss}},\ and\ \citenamefont {Lesanovsky}}]{nill2024avalanche}%
  \BibitemOpen
  \bibfield  {author} {\bibinfo {author} {\bibfnamefont {C.}~\bibnamefont {Nill}}, \bibinfo {author} {\bibfnamefont {A.}~\bibnamefont {Cabot}}, \bibinfo {author} {\bibfnamefont {A.}~\bibnamefont {Trautmann}}, \bibinfo {author} {\bibfnamefont {C.}~\bibnamefont {Gro{\ss}}},\ and\ \bibinfo {author} {\bibfnamefont {I.}~\bibnamefont {Lesanovsky}},\ }\href@noop {} {\bibfield  {journal} {\bibinfo  {journal} {Physical Review Letters}\ }\textbf {\bibinfo {volume} {133}},\ \bibinfo {pages} {073603} (\bibinfo {year} {2024})}\BibitemShut {NoStop}%
\bibitem [{\citenamefont {Magoni}\ \emph {et~al.}(2023)\citenamefont {Magoni}, \citenamefont {Joshi},\ and\ \citenamefont {Lesanovsky}}]{magoni2023molecular}%
  \BibitemOpen
  \bibfield  {author} {\bibinfo {author} {\bibfnamefont {M.}~\bibnamefont {Magoni}}, \bibinfo {author} {\bibfnamefont {R.}~\bibnamefont {Joshi}},\ and\ \bibinfo {author} {\bibfnamefont {I.}~\bibnamefont {Lesanovsky}},\ }\href@noop {} {\bibfield  {journal} {\bibinfo  {journal} {Physical Review Letters}\ }\textbf {\bibinfo {volume} {131}},\ \bibinfo {pages} {093002} (\bibinfo {year} {2023})}\BibitemShut {NoStop}%
\bibitem [{\citenamefont {Magoni}\ \emph {et~al.}(2024)\citenamefont {Magoni}, \citenamefont {Nill},\ and\ \citenamefont {Lesanovsky}}]{magoni2024coherent}%
  \BibitemOpen
  \bibfield  {author} {\bibinfo {author} {\bibfnamefont {M.}~\bibnamefont {Magoni}}, \bibinfo {author} {\bibfnamefont {C.}~\bibnamefont {Nill}},\ and\ \bibinfo {author} {\bibfnamefont {I.}~\bibnamefont {Lesanovsky}},\ }\href@noop {} {\bibfield  {journal} {\bibinfo  {journal} {Physical Review Letters}\ }\textbf {\bibinfo {volume} {132}},\ \bibinfo {pages} {133401} (\bibinfo {year} {2024})}\BibitemShut {NoStop}%
\bibitem [{\citenamefont {Wintermantel}\ \emph {et~al.}(2021)\citenamefont {Wintermantel}, \citenamefont {Buchhold}, \citenamefont {Shevate}, \citenamefont {Morgado}, \citenamefont {Wang}, \citenamefont {Lochead}, \citenamefont {Diehl},\ and\ \citenamefont {Whitlock}}]{wintermantel2021epidemic}%
  \BibitemOpen
  \bibfield  {author} {\bibinfo {author} {\bibfnamefont {T.}~\bibnamefont {Wintermantel}}, \bibinfo {author} {\bibfnamefont {M.}~\bibnamefont {Buchhold}}, \bibinfo {author} {\bibfnamefont {S.}~\bibnamefont {Shevate}}, \bibinfo {author} {\bibfnamefont {M.}~\bibnamefont {Morgado}}, \bibinfo {author} {\bibfnamefont {Y.}~\bibnamefont {Wang}}, \bibinfo {author} {\bibfnamefont {G.}~\bibnamefont {Lochead}}, \bibinfo {author} {\bibfnamefont {S.}~\bibnamefont {Diehl}},\ and\ \bibinfo {author} {\bibfnamefont {S.}~\bibnamefont {Whitlock}},\ }\href@noop {} {\bibfield  {journal} {\bibinfo  {journal} {Nature Communications}\ }\textbf {\bibinfo {volume} {12}},\ \bibinfo {pages} {103} (\bibinfo {year} {2021})}\BibitemShut {NoStop}%
\bibitem [{\citenamefont {Helmrich}\ \emph {et~al.}(2020)\citenamefont {Helmrich}, \citenamefont {Arias}, \citenamefont {Lochead}, \citenamefont {Wintermantel}, \citenamefont {Buchhold}, \citenamefont {Diehl},\ and\ \citenamefont {Whitlock}}]{helmrich2020signatures}%
  \BibitemOpen
  \bibfield  {author} {\bibinfo {author} {\bibfnamefont {S.}~\bibnamefont {Helmrich}}, \bibinfo {author} {\bibfnamefont {A.}~\bibnamefont {Arias}}, \bibinfo {author} {\bibfnamefont {G.}~\bibnamefont {Lochead}}, \bibinfo {author} {\bibfnamefont {T.}~\bibnamefont {Wintermantel}}, \bibinfo {author} {\bibfnamefont {M.}~\bibnamefont {Buchhold}}, \bibinfo {author} {\bibfnamefont {S.}~\bibnamefont {Diehl}},\ and\ \bibinfo {author} {\bibfnamefont {S.}~\bibnamefont {Whitlock}},\ }\href@noop {} {\bibfield  {journal} {\bibinfo  {journal} {Nature}\ }\textbf {\bibinfo {volume} {577}},\ \bibinfo {pages} {481} (\bibinfo {year} {2020})}\BibitemShut {NoStop}%
\bibitem [{\citenamefont {Brady}\ \emph {et~al.}(2024)\citenamefont {Brady}, \citenamefont {Ohler}, \citenamefont {Otterbach},\ and\ \citenamefont {Fleischhauer}}]{fleischhauerpercolation}%
  \BibitemOpen
  \bibfield  {author} {\bibinfo {author} {\bibfnamefont {D.}~\bibnamefont {Brady}}, \bibinfo {author} {\bibfnamefont {S.}~\bibnamefont {Ohler}}, \bibinfo {author} {\bibfnamefont {J.}~\bibnamefont {Otterbach}},\ and\ \bibinfo {author} {\bibfnamefont {M.}~\bibnamefont {Fleischhauer}},\ }\href {https://doi.org/10.1103/PhysRevLett.133.173401} {\bibfield  {journal} {\bibinfo  {journal} {Physical Review Letters}\ }\textbf {\bibinfo {volume} {133}},\ \bibinfo {pages} {173401} (\bibinfo {year} {2024})}\BibitemShut {NoStop}%
\bibitem [{\citenamefont {Urvoy}\ \emph {et~al.}(2015)\citenamefont {Urvoy}, \citenamefont {Ripka}, \citenamefont {Lesanovsky}, \citenamefont {Booth}, \citenamefont {Shaffer}, \citenamefont {Pfau},\ and\ \citenamefont {L{\"o}w}}]{urvoy2015strongly}%
  \BibitemOpen
  \bibfield  {author} {\bibinfo {author} {\bibfnamefont {A.}~\bibnamefont {Urvoy}}, \bibinfo {author} {\bibfnamefont {F.}~\bibnamefont {Ripka}}, \bibinfo {author} {\bibfnamefont {I.}~\bibnamefont {Lesanovsky}}, \bibinfo {author} {\bibfnamefont {D.}~\bibnamefont {Booth}}, \bibinfo {author} {\bibfnamefont {J.}~\bibnamefont {Shaffer}}, \bibinfo {author} {\bibfnamefont {T.}~\bibnamefont {Pfau}},\ and\ \bibinfo {author} {\bibfnamefont {R.}~\bibnamefont {L{\"o}w}},\ }\href@noop {} {\bibfield  {journal} {\bibinfo  {journal} {Physical Review Letters}\ }\textbf {\bibinfo {volume} {114}},\ \bibinfo {pages} {203002} (\bibinfo {year} {2015})}\BibitemShut {NoStop}%
\bibitem [{\citenamefont {Ding}\ \emph {et~al.}(2024)\citenamefont {Ding}, \citenamefont {Bai}, \citenamefont {Liu}, \citenamefont {Shi}, \citenamefont {Guo}, \citenamefont {Li},\ and\ \citenamefont {Adams}}]{ding2024ergodicity}%
  \BibitemOpen
  \bibfield  {author} {\bibinfo {author} {\bibfnamefont {D.}~\bibnamefont {Ding}}, \bibinfo {author} {\bibfnamefont {Z.}~\bibnamefont {Bai}}, \bibinfo {author} {\bibfnamefont {Z.}~\bibnamefont {Liu}}, \bibinfo {author} {\bibfnamefont {B.}~\bibnamefont {Shi}}, \bibinfo {author} {\bibfnamefont {G.}~\bibnamefont {Guo}}, \bibinfo {author} {\bibfnamefont {W.}~\bibnamefont {Li}},\ and\ \bibinfo {author} {\bibfnamefont {C.~S.}\ \bibnamefont {Adams}},\ }\href@noop {} {\bibfield  {journal} {\bibinfo  {journal} {Science Advances}\ }\textbf {\bibinfo {volume} {10}},\ \bibinfo {pages} {eadl5893} (\bibinfo {year} {2024})}\BibitemShut {NoStop}%
\bibitem [{\citenamefont {Wu}\ \emph {et~al.}(2024)\citenamefont {Wu}, \citenamefont {Xie}, \citenamefont {Li}, \citenamefont {Guo}, \citenamefont {Zou},\ and\ \citenamefont {Xiang}}]{wu2024nonlinearity}%
  \BibitemOpen
  \bibfield  {author} {\bibinfo {author} {\bibfnamefont {K.-D.}\ \bibnamefont {Wu}}, \bibinfo {author} {\bibfnamefont {C.}~\bibnamefont {Xie}}, \bibinfo {author} {\bibfnamefont {C.-F.}\ \bibnamefont {Li}}, \bibinfo {author} {\bibfnamefont {G.-C.}\ \bibnamefont {Guo}}, \bibinfo {author} {\bibfnamefont {C.-L.}\ \bibnamefont {Zou}},\ and\ \bibinfo {author} {\bibfnamefont {G.-Y.}\ \bibnamefont {Xiang}},\ }\href@noop {} {\bibfield  {journal} {\bibinfo  {journal} {Science Advances}\ }\textbf {\bibinfo {volume} {10}},\ \bibinfo {pages} {eado8130} (\bibinfo {year} {2024})}\BibitemShut {NoStop}%
\bibitem [{\citenamefont {Liu}\ \emph {et~al.}(2024)\citenamefont {Liu}, \citenamefont {Sun}, \citenamefont {Cabot}, \citenamefont {Carollo}, \citenamefont {Zhang}, \citenamefont {Zhang}, \citenamefont {Zhang}, \citenamefont {Liu}, \citenamefont {Han}, \citenamefont {Li} \emph {et~al.}}]{liu2024microwave}%
  \BibitemOpen
  \bibfield  {author} {\bibinfo {author} {\bibfnamefont {Z.-K.}\ \bibnamefont {Liu}}, \bibinfo {author} {\bibfnamefont {K.-H.}\ \bibnamefont {Sun}}, \bibinfo {author} {\bibfnamefont {A.}~\bibnamefont {Cabot}}, \bibinfo {author} {\bibfnamefont {F.}~\bibnamefont {Carollo}}, \bibinfo {author} {\bibfnamefont {J.}~\bibnamefont {Zhang}}, \bibinfo {author} {\bibfnamefont {Z.-Y.}\ \bibnamefont {Zhang}}, \bibinfo {author} {\bibfnamefont {L.-H.}\ \bibnamefont {Zhang}}, \bibinfo {author} {\bibfnamefont {B.}~\bibnamefont {Liu}}, \bibinfo {author} {\bibfnamefont {T.-Y.}\ \bibnamefont {Han}}, \bibinfo {author} {\bibfnamefont {Q.}~\bibnamefont {Li}}, \emph {et~al.},\ }\href@noop {} {\bibfield  {journal} {\bibinfo  {journal} {arXiv preprint arXiv:2402.04815}\ } (\bibinfo {year} {2024})}\BibitemShut {NoStop}%
\bibitem [{\citenamefont {Johansson}\ \emph {et~al.}(2012)\citenamefont {Johansson}, \citenamefont {Nation},\ and\ \citenamefont {Nori}}]{johansson2012qutip}%
  \BibitemOpen
  \bibfield  {author} {\bibinfo {author} {\bibfnamefont {J.~R.}\ \bibnamefont {Johansson}}, \bibinfo {author} {\bibfnamefont {P.~D.}\ \bibnamefont {Nation}},\ and\ \bibinfo {author} {\bibfnamefont {F.}~\bibnamefont {Nori}},\ }\href@noop {} {\bibfield  {journal} {\bibinfo  {journal} {Computer Physics Communications}\ }\textbf {\bibinfo {volume} {183}},\ \bibinfo {pages} {1760} (\bibinfo {year} {2012})}\BibitemShut {NoStop}%
\bibitem [{\citenamefont {Lambert}\ \emph {et~al.}(2024)\citenamefont {Lambert}, \citenamefont {Gigu{\`e}re}, \citenamefont {Menczel}, \citenamefont {Li}, \citenamefont {Hopf}, \citenamefont {Su{\'a}rez}, \citenamefont {Gali}, \citenamefont {Lishman}, \citenamefont {Gadhvi}, \citenamefont {Agarwal} \emph {et~al.}}]{lambert2024qutip}%
  \BibitemOpen
  \bibfield  {author} {\bibinfo {author} {\bibfnamefont {N.}~\bibnamefont {Lambert}}, \bibinfo {author} {\bibfnamefont {E.}~\bibnamefont {Gigu{\`e}re}}, \bibinfo {author} {\bibfnamefont {P.}~\bibnamefont {Menczel}}, \bibinfo {author} {\bibfnamefont {B.}~\bibnamefont {Li}}, \bibinfo {author} {\bibfnamefont {P.}~\bibnamefont {Hopf}}, \bibinfo {author} {\bibfnamefont {G.}~\bibnamefont {Su{\'a}rez}}, \bibinfo {author} {\bibfnamefont {M.}~\bibnamefont {Gali}}, \bibinfo {author} {\bibfnamefont {J.}~\bibnamefont {Lishman}}, \bibinfo {author} {\bibfnamefont {R.}~\bibnamefont {Gadhvi}}, \bibinfo {author} {\bibfnamefont {R.}~\bibnamefont {Agarwal}}, \emph {et~al.},\ }\href@noop {} {\bibfield  {journal} {\bibinfo  {journal} {arXiv preprint arXiv:2412.04705}\ } (\bibinfo {year} {2024})}\BibitemShut {NoStop}%
\bibitem [{\citenamefont {Bluvstein}\ \emph {et~al.}(2022)\citenamefont {Bluvstein}, \citenamefont {Levine}, \citenamefont {Semeghini}, \citenamefont {Wang}, \citenamefont {Ebadi}, \citenamefont {Kalinowski}, \citenamefont {Keesling}, \citenamefont {Maskara}, \citenamefont {Pichler}, \citenamefont {Greiner} \emph {et~al.}}]{bluvstein2022quantum}%
  \BibitemOpen
  \bibfield  {author} {\bibinfo {author} {\bibfnamefont {D.}~\bibnamefont {Bluvstein}}, \bibinfo {author} {\bibfnamefont {H.}~\bibnamefont {Levine}}, \bibinfo {author} {\bibfnamefont {G.}~\bibnamefont {Semeghini}}, \bibinfo {author} {\bibfnamefont {T.~T.}\ \bibnamefont {Wang}}, \bibinfo {author} {\bibfnamefont {S.}~\bibnamefont {Ebadi}}, \bibinfo {author} {\bibfnamefont {M.}~\bibnamefont {Kalinowski}}, \bibinfo {author} {\bibfnamefont {A.}~\bibnamefont {Keesling}}, \bibinfo {author} {\bibfnamefont {N.}~\bibnamefont {Maskara}}, \bibinfo {author} {\bibfnamefont {H.}~\bibnamefont {Pichler}}, \bibinfo {author} {\bibfnamefont {M.}~\bibnamefont {Greiner}}, \emph {et~al.},\ }\href@noop {} {\bibfield  {journal} {\bibinfo  {journal} {Nature}\ }\textbf {\bibinfo {volume} {604}},\ \bibinfo {pages} {451} (\bibinfo {year} {2022})}\BibitemShut {NoStop}%
\bibitem [{\citenamefont {Evered}\ \emph {et~al.}(2023)\citenamefont {Evered}, \citenamefont {Bluvstein}, \citenamefont {Kalinowski}, \citenamefont {Ebadi}, \citenamefont {Manovitz}, \citenamefont {Zhou}, \citenamefont {Li}, \citenamefont {Geim}, \citenamefont {Wang}, \citenamefont {Maskara} \emph {et~al.}}]{evered2023high}%
  \BibitemOpen
  \bibfield  {author} {\bibinfo {author} {\bibfnamefont {S.~J.}\ \bibnamefont {Evered}}, \bibinfo {author} {\bibfnamefont {D.}~\bibnamefont {Bluvstein}}, \bibinfo {author} {\bibfnamefont {M.}~\bibnamefont {Kalinowski}}, \bibinfo {author} {\bibfnamefont {S.}~\bibnamefont {Ebadi}}, \bibinfo {author} {\bibfnamefont {T.}~\bibnamefont {Manovitz}}, \bibinfo {author} {\bibfnamefont {H.}~\bibnamefont {Zhou}}, \bibinfo {author} {\bibfnamefont {S.~H.}\ \bibnamefont {Li}}, \bibinfo {author} {\bibfnamefont {A.~A.}\ \bibnamefont {Geim}}, \bibinfo {author} {\bibfnamefont {T.~T.}\ \bibnamefont {Wang}}, \bibinfo {author} {\bibfnamefont {N.}~\bibnamefont {Maskara}}, \emph {et~al.},\ }\href@noop {} {\bibfield  {journal} {\bibinfo  {journal} {Nature}\ }\textbf {\bibinfo {volume} {622}},\ \bibinfo {pages} {268} (\bibinfo {year} {2023})}\BibitemShut {NoStop}%
\bibitem [{\citenamefont {Rosi}\ \emph {et~al.}(2018)\citenamefont {Rosi}, \citenamefont {Burchianti}, \citenamefont {Conclave}, \citenamefont {Naik}, \citenamefont {Roati}, \citenamefont {Fort},\ and\ \citenamefont {Minardi}}]{rosi2018lambda}%
  \BibitemOpen
  \bibfield  {author} {\bibinfo {author} {\bibfnamefont {S.}~\bibnamefont {Rosi}}, \bibinfo {author} {\bibfnamefont {A.}~\bibnamefont {Burchianti}}, \bibinfo {author} {\bibfnamefont {S.}~\bibnamefont {Conclave}}, \bibinfo {author} {\bibfnamefont {D.~S.}\ \bibnamefont {Naik}}, \bibinfo {author} {\bibfnamefont {G.}~\bibnamefont {Roati}}, \bibinfo {author} {\bibfnamefont {C.}~\bibnamefont {Fort}},\ and\ \bibinfo {author} {\bibfnamefont {F.}~\bibnamefont {Minardi}},\ }\href@noop {} {\bibfield  {journal} {\bibinfo  {journal} {Scientific reports}\ }\textbf {\bibinfo {volume} {8}},\ \bibinfo {pages} {1301} (\bibinfo {year} {2018})}\BibitemShut {NoStop}%
\bibitem [{\citenamefont {Lorenz}\ \emph {et~al.}(2021)\citenamefont {Lorenz}, \citenamefont {Festa}, \citenamefont {Steinert},\ and\ \citenamefont {Gross}}]{10.21468/SciPostPhys.10.3.052}%
  \BibitemOpen
  \bibfield  {author} {\bibinfo {author} {\bibfnamefont {N.}~\bibnamefont {Lorenz}}, \bibinfo {author} {\bibfnamefont {L.}~\bibnamefont {Festa}}, \bibinfo {author} {\bibfnamefont {L.-M.}\ \bibnamefont {Steinert}},\ and\ \bibinfo {author} {\bibfnamefont {C.}~\bibnamefont {Gross}},\ }\href {https://doi.org/10.21468/SciPostPhys.10.3.052} {\bibfield  {journal} {\bibinfo  {journal} {SciPost Phys.}\ }\textbf {\bibinfo {volume} {10}},\ \bibinfo {pages} {052} (\bibinfo {year} {2021})}\BibitemShut {NoStop}%
\bibitem [{\citenamefont {Panja}\ \emph {et~al.}(2024)\citenamefont {Panja}, \citenamefont {Wang}, \citenamefont {Wang}, \citenamefont {Wang}, \citenamefont {Subhankar},\ and\ \citenamefont {Liang}}]{panja2024electric}%
  \BibitemOpen
  \bibfield  {author} {\bibinfo {author} {\bibfnamefont {A.}~\bibnamefont {Panja}}, \bibinfo {author} {\bibfnamefont {Y.}~\bibnamefont {Wang}}, \bibinfo {author} {\bibfnamefont {X.}~\bibnamefont {Wang}}, \bibinfo {author} {\bibfnamefont {J.}~\bibnamefont {Wang}}, \bibinfo {author} {\bibfnamefont {S.}~\bibnamefont {Subhankar}},\ and\ \bibinfo {author} {\bibfnamefont {Q.-Y.}\ \bibnamefont {Liang}},\ }\href@noop {} {\bibfield  {journal} {\bibinfo  {journal} {AIP Advances}\ }\textbf {\bibinfo {volume} {14}} (\bibinfo {year} {2024})}\BibitemShut {NoStop}%
\bibitem [{\citenamefont {Anand}\ \emph {et~al.}(2024)\citenamefont {Anand}, \citenamefont {Bradley}, \citenamefont {White}, \citenamefont {Ramesh}, \citenamefont {Singh},\ and\ \citenamefont {Bernien}}]{Ananddualspeciesrydberg}%
  \BibitemOpen
  \bibfield  {author} {\bibinfo {author} {\bibfnamefont {S.}~\bibnamefont {Anand}}, \bibinfo {author} {\bibfnamefont {C.~E.}\ \bibnamefont {Bradley}}, \bibinfo {author} {\bibfnamefont {R.}~\bibnamefont {White}}, \bibinfo {author} {\bibfnamefont {V.}~\bibnamefont {Ramesh}}, \bibinfo {author} {\bibfnamefont {K.}~\bibnamefont {Singh}},\ and\ \bibinfo {author} {\bibfnamefont {H.}~\bibnamefont {Bernien}},\ }\href {https://doi.org/10.1038/s41567-024-02638-2} {\bibfield  {journal} {\bibinfo  {journal} {Nature Physics}\ }\textbf {\bibinfo {volume} {20}},\ \bibinfo {pages} {1744–1750} (\bibinfo {year} {2024})}\BibitemShut {NoStop}%
\bibitem [{\citenamefont {Wilson}\ \emph {et~al.}(2022)\citenamefont {Wilson}, \citenamefont {Saskin}, \citenamefont {Meng}, \citenamefont {Ma}, \citenamefont {Dilip}, \citenamefont {Burgers},\ and\ \citenamefont {Thompson}}]{wilson2022trapping}%
  \BibitemOpen
  \bibfield  {author} {\bibinfo {author} {\bibfnamefont {J.}~\bibnamefont {Wilson}}, \bibinfo {author} {\bibfnamefont {S.}~\bibnamefont {Saskin}}, \bibinfo {author} {\bibfnamefont {Y.}~\bibnamefont {Meng}}, \bibinfo {author} {\bibfnamefont {S.}~\bibnamefont {Ma}}, \bibinfo {author} {\bibfnamefont {R.}~\bibnamefont {Dilip}}, \bibinfo {author} {\bibfnamefont {A.}~\bibnamefont {Burgers}},\ and\ \bibinfo {author} {\bibfnamefont {J.}~\bibnamefont {Thompson}},\ }\href@noop {} {\bibfield  {journal} {\bibinfo  {journal} {Physical Review Letters}\ }\textbf {\bibinfo {volume} {128}},\ \bibinfo {pages} {033201} (\bibinfo {year} {2022})}\BibitemShut {NoStop}%
\bibitem [{\citenamefont {Bluvstein}\ \emph {et~al.}(2024)\citenamefont {Bluvstein}, \citenamefont {Evered}, \citenamefont {Geim}, \citenamefont {Li}, \citenamefont {Zhou}, \citenamefont {Manovitz}, \citenamefont {Ebadi}, \citenamefont {Cain}, \citenamefont {Kalinowski}, \citenamefont {Hangleiter} \emph {et~al.}}]{bluvstein2024logical}%
  \BibitemOpen
  \bibfield  {author} {\bibinfo {author} {\bibfnamefont {D.}~\bibnamefont {Bluvstein}}, \bibinfo {author} {\bibfnamefont {S.~J.}\ \bibnamefont {Evered}}, \bibinfo {author} {\bibfnamefont {A.~A.}\ \bibnamefont {Geim}}, \bibinfo {author} {\bibfnamefont {S.~H.}\ \bibnamefont {Li}}, \bibinfo {author} {\bibfnamefont {H.}~\bibnamefont {Zhou}}, \bibinfo {author} {\bibfnamefont {T.}~\bibnamefont {Manovitz}}, \bibinfo {author} {\bibfnamefont {S.}~\bibnamefont {Ebadi}}, \bibinfo {author} {\bibfnamefont {M.}~\bibnamefont {Cain}}, \bibinfo {author} {\bibfnamefont {M.}~\bibnamefont {Kalinowski}}, \bibinfo {author} {\bibfnamefont {D.}~\bibnamefont {Hangleiter}}, \emph {et~al.},\ }\href@noop {} {\bibfield  {journal} {\bibinfo  {journal} {Nature}\ }\textbf {\bibinfo {volume} {626}},\ \bibinfo {pages} {58} (\bibinfo {year} {2024})}\BibitemShut {NoStop}%
\bibitem [{\citenamefont {Graham}\ \emph {et~al.}(2022)\citenamefont {Graham}, \citenamefont {Song}, \citenamefont {Scott}, \citenamefont {Poole}, \citenamefont {Phuttitarn}, \citenamefont {Jooya}, \citenamefont {Eichler}, \citenamefont {Jiang}, \citenamefont {Marra}, \citenamefont {Grinkemeyer} \emph {et~al.}}]{graham2022multi}%
  \BibitemOpen
  \bibfield  {author} {\bibinfo {author} {\bibfnamefont {T.}~\bibnamefont {Graham}}, \bibinfo {author} {\bibfnamefont {Y.}~\bibnamefont {Song}}, \bibinfo {author} {\bibfnamefont {J.}~\bibnamefont {Scott}}, \bibinfo {author} {\bibfnamefont {C.}~\bibnamefont {Poole}}, \bibinfo {author} {\bibfnamefont {L.}~\bibnamefont {Phuttitarn}}, \bibinfo {author} {\bibfnamefont {K.}~\bibnamefont {Jooya}}, \bibinfo {author} {\bibfnamefont {P.}~\bibnamefont {Eichler}}, \bibinfo {author} {\bibfnamefont {X.}~\bibnamefont {Jiang}}, \bibinfo {author} {\bibfnamefont {A.}~\bibnamefont {Marra}}, \bibinfo {author} {\bibfnamefont {B.}~\bibnamefont {Grinkemeyer}}, \emph {et~al.},\ }\href@noop {} {\bibfield  {journal} {\bibinfo  {journal} {Nature}\ }\textbf {\bibinfo {volume} {604}},\ \bibinfo {pages} {457} (\bibinfo {year} {2022})}\BibitemShut {NoStop}%
\bibitem [{\citenamefont {Omran}\ \emph {et~al.}(2019)\citenamefont {Omran}, \citenamefont {Levine}, \citenamefont {Keesling}, \citenamefont {Semeghini}, \citenamefont {Wang}, \citenamefont {Ebadi}, \citenamefont {Bernien}, \citenamefont {Zibrov}, \citenamefont {Pichler}, \citenamefont {Choi} \emph {et~al.}}]{omran2019generation}%
  \BibitemOpen
  \bibfield  {author} {\bibinfo {author} {\bibfnamefont {A.}~\bibnamefont {Omran}}, \bibinfo {author} {\bibfnamefont {H.}~\bibnamefont {Levine}}, \bibinfo {author} {\bibfnamefont {A.}~\bibnamefont {Keesling}}, \bibinfo {author} {\bibfnamefont {G.}~\bibnamefont {Semeghini}}, \bibinfo {author} {\bibfnamefont {T.~T.}\ \bibnamefont {Wang}}, \bibinfo {author} {\bibfnamefont {S.}~\bibnamefont {Ebadi}}, \bibinfo {author} {\bibfnamefont {H.}~\bibnamefont {Bernien}}, \bibinfo {author} {\bibfnamefont {A.~S.}\ \bibnamefont {Zibrov}}, \bibinfo {author} {\bibfnamefont {H.}~\bibnamefont {Pichler}}, \bibinfo {author} {\bibfnamefont {S.}~\bibnamefont {Choi}}, \emph {et~al.},\ }\href@noop {} {\bibfield  {journal} {\bibinfo  {journal} {Science}\ }\textbf {\bibinfo {volume} {365}},\ \bibinfo {pages} {570} (\bibinfo {year} {2019})}\BibitemShut {NoStop}%
\bibitem [{Note1()}]{Note1}%
  \BibitemOpen
  \bibinfo {note} {The array can be repositioned to locations that satisfy facilitation conditions required by our transport protocol after the completion of a global excitation pulse.}\BibitemShut {Stop}%
\bibitem [{\citenamefont {Madjarov}\ \emph {et~al.}(2020)\citenamefont {Madjarov}, \citenamefont {Covey}, \citenamefont {Shaw}, \citenamefont {Choi}, \citenamefont {Kale}, \citenamefont {Cooper}, \citenamefont {Pichler}, \citenamefont {Schkolnik}, \citenamefont {Williams},\ and\ \citenamefont {Endres}}]{madjarov2020high}%
  \BibitemOpen
  \bibfield  {author} {\bibinfo {author} {\bibfnamefont {I.~S.}\ \bibnamefont {Madjarov}}, \bibinfo {author} {\bibfnamefont {J.~P.}\ \bibnamefont {Covey}}, \bibinfo {author} {\bibfnamefont {A.~L.}\ \bibnamefont {Shaw}}, \bibinfo {author} {\bibfnamefont {J.}~\bibnamefont {Choi}}, \bibinfo {author} {\bibfnamefont {A.}~\bibnamefont {Kale}}, \bibinfo {author} {\bibfnamefont {A.}~\bibnamefont {Cooper}}, \bibinfo {author} {\bibfnamefont {H.}~\bibnamefont {Pichler}}, \bibinfo {author} {\bibfnamefont {V.}~\bibnamefont {Schkolnik}}, \bibinfo {author} {\bibfnamefont {J.~R.}\ \bibnamefont {Williams}},\ and\ \bibinfo {author} {\bibfnamefont {M.}~\bibnamefont {Endres}},\ }\href@noop {} {\bibfield  {journal} {\bibinfo  {journal} {Nature Physics}\ }\textbf {\bibinfo {volume} {16}},\ \bibinfo {pages} {857} (\bibinfo {year} {2020})}\BibitemShut {NoStop}%
\bibitem [{\citenamefont {Levine}\ \emph {et~al.}(2019)\citenamefont {Levine}, \citenamefont {Keesling}, \citenamefont {Semeghini}, \citenamefont {Omran}, \citenamefont {Wang}, \citenamefont {Ebadi}, \citenamefont {Bernien}, \citenamefont {Greiner}, \citenamefont {Vuleti\ifmmode~\acute{c}\else \'{c}\fi{}}, \citenamefont {Pichler},\ and\ \citenamefont {Lukin}}]{PhysRevLett.123.170503}%
  \BibitemOpen
  \bibfield  {author} {\bibinfo {author} {\bibfnamefont {H.}~\bibnamefont {Levine}}, \bibinfo {author} {\bibfnamefont {A.}~\bibnamefont {Keesling}}, \bibinfo {author} {\bibfnamefont {G.}~\bibnamefont {Semeghini}}, \bibinfo {author} {\bibfnamefont {A.}~\bibnamefont {Omran}}, \bibinfo {author} {\bibfnamefont {T.~T.}\ \bibnamefont {Wang}}, \bibinfo {author} {\bibfnamefont {S.}~\bibnamefont {Ebadi}}, \bibinfo {author} {\bibfnamefont {H.}~\bibnamefont {Bernien}}, \bibinfo {author} {\bibfnamefont {M.}~\bibnamefont {Greiner}}, \bibinfo {author} {\bibfnamefont {V.}~\bibnamefont {Vuleti\ifmmode~\acute{c}\else \'{c}\fi{}}}, \bibinfo {author} {\bibfnamefont {H.}~\bibnamefont {Pichler}},\ and\ \bibinfo {author} {\bibfnamefont {M.~D.}\ \bibnamefont {Lukin}},\ }\href {https://doi.org/10.1103/PhysRevLett.123.170503} {\bibfield  {journal} {\bibinfo  {journal} {Physical Review Letters}\ }\textbf {\bibinfo {volume} {123}},\ \bibinfo {pages} {170503} (\bibinfo {year} {2019})}\BibitemShut {NoStop}%
\bibitem [{\citenamefont {Ma}\ \emph {et~al.}(2023)\citenamefont {Ma}, \citenamefont {Liu}, \citenamefont {Peng}, \citenamefont {Zhang}, \citenamefont {Jandura}, \citenamefont {Claes}, \citenamefont {Burgers}, \citenamefont {Pupillo}, \citenamefont {Puri},\ and\ \citenamefont {Thompson}}]{thompsongate}%
  \BibitemOpen
  \bibfield  {author} {\bibinfo {author} {\bibfnamefont {S.}~\bibnamefont {Ma}}, \bibinfo {author} {\bibfnamefont {G.}~\bibnamefont {Liu}}, \bibinfo {author} {\bibfnamefont {P.}~\bibnamefont {Peng}}, \bibinfo {author} {\bibfnamefont {B.}~\bibnamefont {Zhang}}, \bibinfo {author} {\bibfnamefont {S.}~\bibnamefont {Jandura}}, \bibinfo {author} {\bibfnamefont {J.}~\bibnamefont {Claes}}, \bibinfo {author} {\bibfnamefont {A.~P.}\ \bibnamefont {Burgers}}, \bibinfo {author} {\bibfnamefont {G.}~\bibnamefont {Pupillo}}, \bibinfo {author} {\bibfnamefont {S.}~\bibnamefont {Puri}},\ and\ \bibinfo {author} {\bibfnamefont {J.~D.}\ \bibnamefont {Thompson}},\ }\href {https://doi.org/10.1038/s41586-023-06438-1} {\bibfield  {journal} {\bibinfo  {journal} {Nature}\ }\textbf {\bibinfo {volume} {622}},\ \bibinfo {pages} {279–284} (\bibinfo {year} {2023})}\BibitemShut {NoStop}%
\end{thebibliography}%

\end{document}